\documentstyle[12pt,psfig,epsfig]{article}

\newcommand{\beq}{\begin{equation}}
\newcommand{\eeq}{\end{equation}}
\newcommand{\be}{\begin{eqnarray}}
\newcommand{\ee}{\end{eqnarray}}

\def\kl{{KamLAND~}}

\begin{document}

\begin{flushright}
SINP/TNP/02-35\\
SISSA 92/2002/EP\
\end{flushright}

\vskip 5pt

\begin{center} 
{\bf The Solar Neutrino Problem after the first results from Kamland 
}
 
\vskip 5pt
{Abhijit Bandyopadhyay$^a$\footnote{email: abhi@theory.saha.ernet.in},
Sandhya Choubey$^b$\footnote{email: sandhya@he.sissa.it},
Raj Gandhi$^c$\footnote{email: raj@mri.ernet.in},
\\
Srubabati Goswami$^c$\footnote{email: sruba@mri.ernet.in},
D.P. Roy$^d$\footnote{email: dproy@theory.tifr.res.in}
\\
$^a${\it Saha Institute of Nuclear Physics,1/AF, Bidhannagar,
Kolkata 700 064, India}\\
$^b${\it INFN, Sezione di Trieste and 
Scuola Internazionale Superiore di Studi Avanzati,\\
I-34014, 
Trieste, Italy}\\
$^c${\it Harish-Chandra Research Institute, Chhatnag Road, Jhusi,\\
Allahabad - 211-019, India}\\
$^d${\it Tata Institute of Fundamental Research,
Homi Bhabha Road, Mumbai 400005, India}
}
\end{center}

\begin{abstract}
The first results from the KamLAND experiment  
have provided confirmational evidence for
 the Large Mixing Angle (LMA)   
Mikheyev-Smirnov-Wolfenstein (MSW) 
solution
to the solar neutrino problem.  
We do a global analysis of solar and the recently announced
KamLAND data (both rate and spectrum) and investigate 
its effect on the allowed region in the $\Delta m^2-\tan^2\theta$ 
plane. 
The best-fit from a  combined analysis which uses the KamLAND rate 
plus global solar data comes 
at $\Delta m^2 = 6.06 \times 10^{-5}$ eV $^2$ and $\tan^2\theta=0.42
$, very close to the global solar best-fit, leaving a large allowed region 
within the global solar LMA contour.
The inclusion of the KamLAND spectral data in the 
global fit gives 
a best-fit $\Delta m^2 = 7.17 \times 10^{-5}$ eV $^2$ and $\tan^2\theta=0.43
$ and constrains the allowed areas within LMA,
leaving essentially two allowed 
zones. Maximal mixing though allowed by the \kl data alone is 
disfavored by the global solar data and remains disallowed at about 
$3\sigma$. 
The low $\Delta m^2$ solution (LOW)
is now ruled out at about 5$\sigma$ with respect to 
the LMA solution.  
\end{abstract}

\vskip 10pt
\section{Introduction}
It is fair to say that the recently  announced first results  of  the
Kamioka Liquid scintillator 
Anti-Neutrino Detector (KamLAND) experiment \cite{kam_1}
constitute a highly anticipated  milestone in our
understanding and resolution
of the three decade old solar neutrino problem.
The origins of this puzzle lie in the  early deficit measurements of the
solar neutrino flux in the pioneering Homestake chlorine experiment\cite{cl}. 
This discrepancy between the expected rate, as predicted by increasingly 
refined solar model calculations
\cite{jnb_1} and the measured one has been  subsequently
confirmed and buttressed over the years by results from the 
$^{71}{Ga}$ experiments SAGE, GALLEX and GNO \cite{gno,sage},
the  Kamiokande and  the Super-Kamiokande experiments (SK) 
\cite{superk},
and most recently from the Sudbury
Neutrino Observatory (SNO) \cite{sno1,sno2}. In particular, 
SK has provided valuable  zenith angle and energy spectrum 
information  in
addition to total rate measurements of the high energy Boron
flux, and SNO has provided crucial neutral current 
(NC) and charged current (CC) rate data along with spectrum  results. 
Over the years, these experimental results have
been culled together with our understanding of neutrino mass, 
mixing and resonant matter oscillations  to obtain 
the allowed parameter space in terms of the mixing angle
$\tan^2\theta$
and mass-squared difference $\Delta m^2$  of the neutrino states. 
The analysis of global solar data carried out by various groups
favours  the 
LMA solution based on  MSW resonant matter 
oscillations \cite{msw}
 as the most probable resolution of the solar neutrino
problem
\cite{a1,Bandyopadhyay:2002xj,Choubey:2002nc,Bandyopadhyay:2002qg,
Bahcall:2002hv,a4,a5,a6,Strumia:2002rv,Fogli:2002pt,a9}. 

KamLAND \cite{kam_2} is a 1 kton liquid scintillator neutrino
detector,
designed specifically to
test the LMA solution. 
It is 
located at the earlier Kamiokande
site in the Kamioka mine in Japan.
Its  main objective is to look for oscillation of $\bar{\nu_e}$
coming from
Japanese nuclear power reactors
situated at distances ranging from $\sim$ $80$ km  to $800$ km.
The bulk ($\sim 79\%$)
of the measured flux is however from reactors which are at distances between
138 km to 214 km.
The $\bar\nu_e$s are detected
via the inverse beta decay reaction
$\bar{\nu}_e+p\rightarrow e^{+}+n$. Both the
scintillation emitted by the positron as it
moves through the detection medium,
and its subsequent annihilation with an electron are recorded.
The delayed coincidence of the positron with
the  2.2 MeV $\gamma$-ray from the capture of the neutron
constitutes  a largely background free  signal.
The total visible energy ($E_{vis}$) corresponds to
$E_{e^+}+m_{e}$, where $E_{e^+}$ is the 
total energy of the positron and
$m_e$ the electron mass.
The positron energy
is related to the incoming antineutrino energy as
$E_{e^+}=E_{\nu}- {\bar E_{rec}} -(m_n -  m_p)$~MeV 
($m_n - m_p = 1.293~$MeV is the
neutron--proton mass difference). 
${\bar E_{rec}}$ is the average neutron
recoil energy calculated here  using \cite{beavo}.
The energy resolution is
$\sigma(E)/E=7.5\%/\sqrt{E}$, $E$ is in MeV. 

The first data from  KamLAND 
gives the ratio of the observed number of events
to the expected number of events to be \cite{kam_1}
\be
R_{KL}=0.611 \pm 0.085 (stat) \pm 0.041 (syst)
\label{data}
\ee
for an exposure of
162 ton-yr and   
a visible energy
above 2.6 MeV  
\footnote{Below this energy the background due to the geophysical
neutrinos dominate.}.
They have also presented the observed positron energy spectrum. 

In a pre-\kl  analysis \cite{raj} we have shown that an energy 
integrated rate in the range 0.3 -0.8 will provide confirmation for the 
LMA solution. In particular the solar LMA best-fit predicted 
a \kl rate of 0.65 which is close to the observed rate. 
This is  the first confirmation 
of the  
LMA solution to the solar neutrino problem
using terrestrial neutrino sources. 
We also showed that 
for a rate below 0.9 
the LOW solution to the solar neutrino problem 
is disallowed at more than 3$\sigma$. 
Hence in this paper we focus on the LMA solution
and perform a global analysis which combines\\ 
(i)KamLAND rate and  global solar data \\
(ii)KamLAND spectrum and global solar data\\
We  find the  
allowed area  from each of the above analyses 
and discuss  their contributions in sharpening 
our knowledge of  neutrino mass and mixing parameters.  
The current KamLAND and global solar data split the allowed 
LMA region in two parts 
 -- a low $\Delta m^2$ region (low-LMA) and 
a high $\Delta m^2$ region (high-LMA), which has less
(by $\approx 2\sigma$) statistical 
significance. 
A more 
precise determination of $\Delta m^2$ and 
$\tan^2\theta$ should be possible with increased statistics 
and reduced systematics of the spectral data from KamLAND  
\cite{bar,mur,deg,str,gg,aliani,Fogli:2002pb,raj}.  
We demonstrate the potential of 1 kton-yr spectral data 
in discriminating between the two allowed regions and
further constraining the parameter values by
simulating the spectrum at different values of $\Delta m^2$ and 
$\tan^2\theta$ selected from the allowed area of the  
global solar+KamLAND  analysis.  
We find that 
if the true spectrum corresponds to that
simulated 
at points in the low-LMA region then with 
1 kton-yr exposure the high-LMA part 
can be further disfavoured. 
For spectrum simulated at high-LMA values however 
the ambiguity between the two zones 
persists. 

\section{Analysis and Results}



The total event-rate/sec in the KamLAND detector is given as \cite{raj}
\begin{eqnarray}
N_{KL} &= &
\int dE_{\nu}\sigma(E_{\nu}) N_p \sum_{i} S_i \frac{P_i(\bar{\nu}_e
\leftrightarrow\bar{\nu}_e)}{4\pi d_i^2}
\label{eventrate}
\end{eqnarray}
where $\sigma({E_{\nu}})$ denotes the cross-section; 
$S_{i}$ denotes the spectrum from a given reactor $i$ and involves the 
neutrino spectrum from the fission of a 
particular isotope, the characteristic energy released per fission 
by the isotope and the 
fractional abundance of the isotopes.  
For further details of the spectrum, cross-section, fuel composition 
etc. we refer the reader to \cite{raj}. 
$N_p$ denotes the number of target protons. 
The declared KamLAND data corresponds to 
a fiducial mass of  408 ton,
resulting in 3.46 $\times$ 10$^{31}$ free target protons\cite{kam_1}. 
Relative fission yields for the various fuel
isotopes are also taken in accordance with  \cite{kam_1},
as is the integrated thermal power flux of 254 Joule/cm$^2$.
$P_i(\bar{\nu}_e\leftrightarrow\bar{\nu}_e)$ is 
the two-generation
survival probability for the antineutrinos from each of the reactors
$i$ and
$d_i$ is the distance of reactor $i$ to KamLAND in $km$.
In addition,  we include  the 
event selection criteria used by the KamLAND collaboration
corresponding to an efficiency of $78.3\%$ \cite{kam_1}. 

We first do a statistical
analysis of the KamLAND rate and global solar data. 
For KamLAND rate we  
define the $\chi^2$ as
\beq
\chi^2_{KL} = \frac{(R_{KL}^{expt} - R_{KL}^{theory})^2}{\sigma^2}
\label{chir}
\eeq
where $\sigma = \sqrt{\sigma_{syst}^2 + \sigma_{stat}^2}$,
$\sigma_{syst}$ and $\sigma_{stat}$ being the total systematic and
statistical error in the KamLAND data respectively (cf. Eq.(\ref{data})).
\be
R_{KL} = \frac{N_{KL}}{N^0_{KL}}
\ee
$N^0_{KL}$ is obtained from  
Eq.(\ref{eventrate}) with $P(\bar{\nu}_e
\leftrightarrow\bar{\nu}_e)=1$.

For the solar analysis we define the 
$\chi^2$ function in the ``covariance'' approach as
\be
\chi^2_{\odot} = \sum_{i,j=1}^N (R_i^{\rm expt}-R_i^{\rm theory})
(\sigma_{ij}^2)^{-1}(R_j^{\rm expt}-R_j^{\rm theory})
\label{chi2}
\ee
where $R_{i}$ are the solar data points, $N$ is
the number of data points (80 in our case) and
$(\sigma_{ij}^2)^{-1}$ is the inverse of the covariance matrix,
containing the squares of the correlated and uncorrelated experimental
and theoretical errors.
We use the
data on total rate from the Cl experiment, the
combined rate from the Ga experiments (SAGE+GALLEX+GNO),
the 1496 day data on the SK zenith angle energy spectrum and
the combined SNO day-night spectrum.
For further details of our solar analysis we refer the reader
to \cite{Choubey:2002nc,Bandyopadhyay:2002qg}.

\begin{figure}[t]
\centerline{\psfig{figure=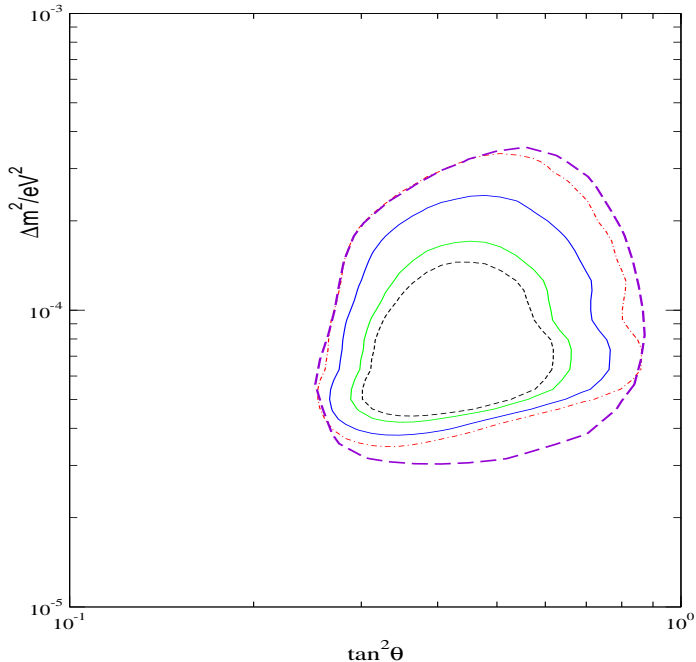,height=4.in,width=4.in}}
\caption{The 90\%, 95\%, 99\% and 99.73\%($3\sigma$) C.L. contours
from a $\chi^2$ analysis using  
KamLAND rate + global solar data.
The dashed line shows the presently allowed  
$3\sigma$ solar contour.}
\label{solkamrate}
\end{figure}

\begin{figure}[t]
\centerline{\psfig{figure= 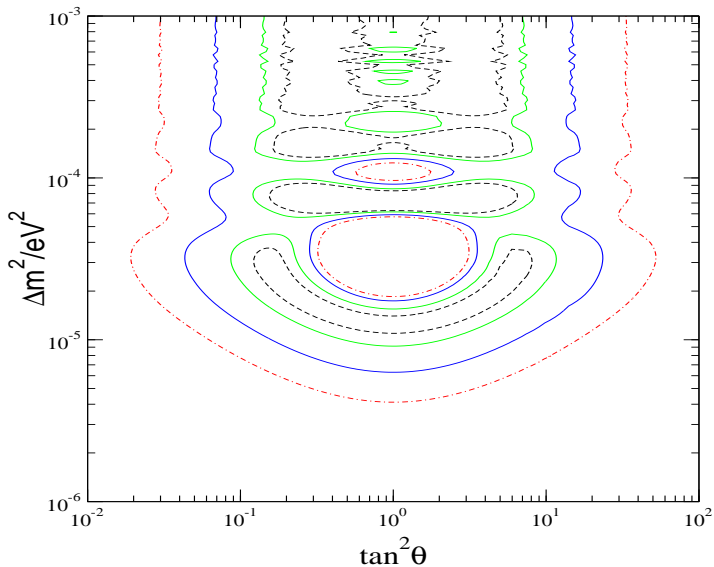,height=4.5in,width=4.in}}
\caption{The 90\%, 95\%, 99\% and 99.73\%($3\sigma$) C.L. contours from 
a $\chi^2$ analysis using  
the 
KamLAND  spectrum data. 
}
\label{rateshape}
\end{figure}

The $\chi^2$ for the combined solar and KamLAND rate analysis is
defined as 
\beq
\chi^2 = \chi^2_{\odot} + \chi^2_{KL}
\label{chi1}
\eeq
The best-fit after including the KamLAND data comes at 
$\Delta m^2 = 6.06 \times 10^{-5}$ eV$^2$ and $\tan^2\theta = 0.42$.
Thus the best-fit point does not change significantly with respect to 
that obtained from only solar analysis \cite{Choubey:2002nc}.
In Figure \ref{solkamrate}
we draw the 90\%, 95\%, 99\% and 99.73\% C.L.
allowed area in the LMA region from a combined
solar+KamLAND rate analysis.
Superimposed on that we show the $3\sigma$ (99.73\% C.L.) 
allowed area from
solar data alone.
Large area within the LMA regions  is seen to remain allowed. 

\begin{figure}[t]
\centerline{\psfig{figure=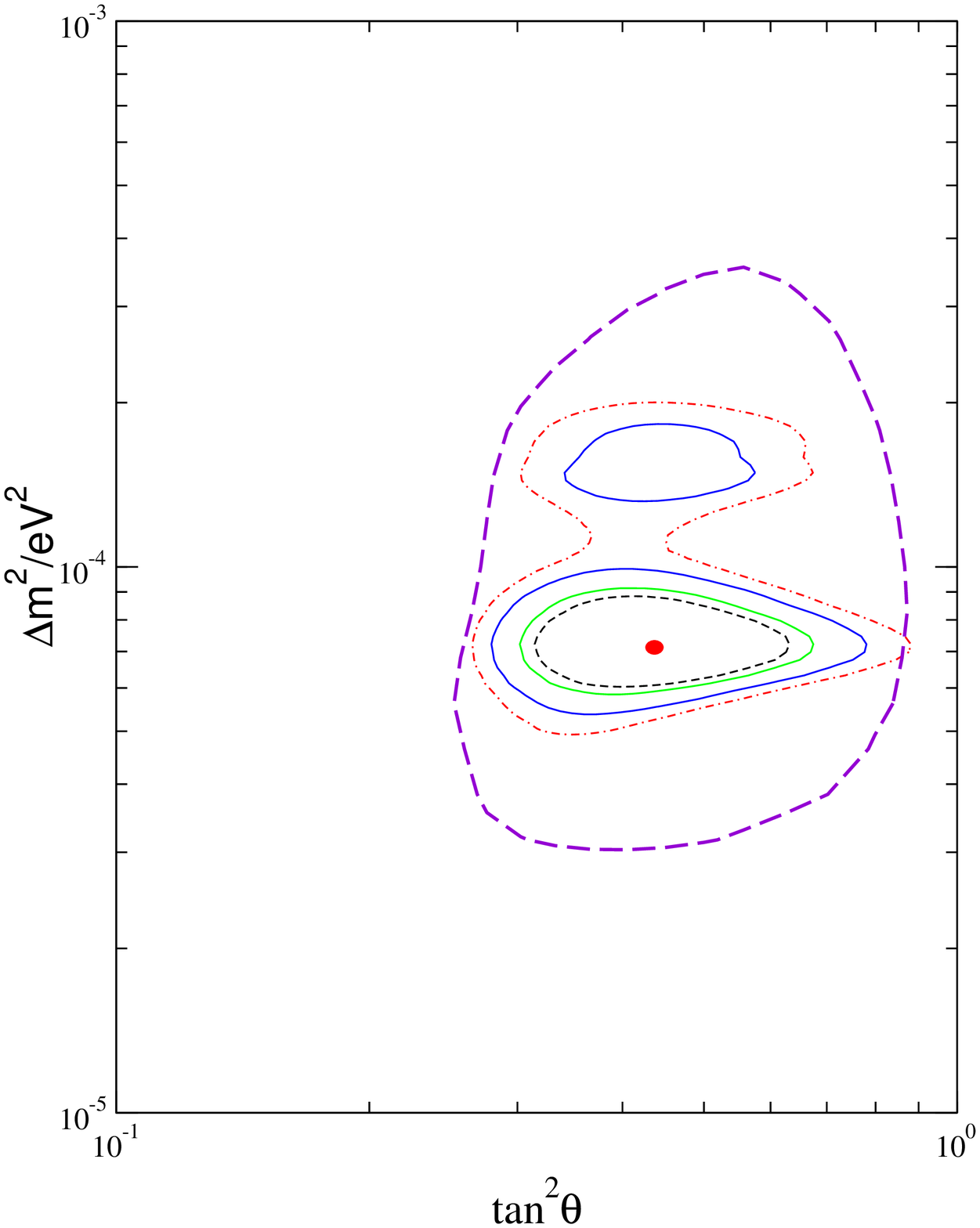,height=4.in,width=4.in}}
\caption{The 90\%, 95\%, 99\% and 99.73\% C.L. contours
from a $\chi^2$ analysis using  
KamLAND spectrum data along with global solar data.
The dashed line shows the presently allowed  
$3\sigma$ only solar contour.}
\label{klrtspecsol}
\end{figure}

\begin{table}
\begin{center}
\begin{tabular}{ccccc}
\hline\hline
Data & $\Delta m^2$ &
$\tan^2\theta$&$\chi^2_{min}$\\
Used& in eV$^2$&  & \\
\hline\hline
&$7.17 \times 10^{-5}$&$0.64$ & 5.71\\
\raisebox{1.5ex}[0pt]\kl&
$1.50\times 10^{-4}$ & 0.34 & 8.24\\
\hline\hline
&$7.17\times 10^{-5}$& 0.43 & 74.39 \\
\raisebox{1.5ex}[0pt]{\kl + Solar} & 
$1.48\times 10^{-4}$& 0.44 & 81.51  \\
\hline\hline
\end{tabular}
\caption
{The $\chi^2_{min}$
and the best-fit values of the oscillation
parameters obtained from the analysis of the \kl spectrum data alone and
from the global analysis of the \kl spectrum data and the solar neutrino
data. 
}
\label{bftab}
\end{center}
\end{table}

Apart from the data on energy integrated total rates, 
KamLAND collaboration has also provided the 
observed positron visible energy spectrum, albeit with 
low statistics.  
We incorporate this in our analysis 
to extract the shape information from this data 
as the  spectral distortion is a very sensitive probe
of $\Delta m^2$.
For the \kl spectral data, we perform our analysis 
using a definition of $\chi^2_{klspec}$ assuming the data to be 
Poisson-distributed which is appropriate for data with low 
statistics. For this case
\be
\chi^2_{klspec}=
\sum_{i}\left[2(X_n S_{KL,i}^{theory} - S_{KL,i}^{expt}) 
+ 2 S_{KL,i}^{expt} \ln(\frac{S_{KL,i}^{expt}}
{X_n S_{KL,i}^{theory}})\right] + \frac{(X_n -1)^2}{\sigma^2_{sys}}
\label{chip}
\ee
where $\sigma_{sys}$ is taken to be 6.42\% \cite{kam_1} and a  
normalisation factor $X_n$ is allowed to vary freely.
The sum is over the 13 KamLAND spectral bins.
In \cite{kam_1} the errors  for the shape distortion are attributed
to
energy scale, energy resolution, $\bar\nu_e$ spectrum and fiducial volume.
A more refined statistical analysis would involve evaluating the
systematic
errors in each bin due to these sources at each
$\Delta m^2$ and $\tan^2\theta$ as well as taking into account of the
background events and their errors in each bin.
This will be possible as and when more
detailed information
will be available.

For the spectrum analysis we get the best-fit values of $\Delta m^2$ 
and $\tan^2 \theta$ to be $7.17 \times 10^{-5}$ eV$^2$ and 
0.64 respectively. 
This is close to  
that
obtained by the \kl 
collaboration  but our best-fit $\theta$ is not
maximal as in \cite{kam_1}. 
Apart from the above there are 
other minimas with reduced statistical significance.
In Table 1 we present the best-fit values and 
$\chi^2_{min}$ for the global minima and 
the second minima which is obtained 
at a higher $\Delta m^2$ value. 
In Figure \ref{rateshape} we present the allowed areas in 
$\Delta m^2  - \tan^2\theta$ plane from 
\kl  spectrum analysis. 
This is understandably slightly different from what \kl has obtained 
in \cite{kam_1}.
The KamLAND data analysis procedure as outlined in \cite{kam_1} 
is somewhat different and the full 
details are not known to us. 

Next we do a combined analysis of \kl spectral data
together with 
the global solar data. The $\chi^2$ for the combined analysis is 
defined as the sum of the individual 
contributions. 
We present the results in Table 1. 
The best-fit comes at $\Delta m^2 = 7.17 \times 10^{-5} $ eV$^2$
and $\tan^2\theta = 0.43$. 
From Table 1 we also see that there is a second minima
at a higher $\Delta m^2$ with a reduced statistical
significance by  about 2$\sigma$ with respect to (w.r.t.)
the global minima. 

In Figure \ref{klrtspecsol}
we show the combined allowed area with the solar 
and \kl spectrum analysis 
superimposed on the 3$\sigma$ solar contour. 
The allowed area is seen to be much constricted with the 
inclusion of the \kl spectral data. 
At 90\% C.L. only a small region about the best-fit point remains
allowed.
At 99\% C.L. however there are  two distinct allowed zones  --
one around the global best-fit point (low-LMA) 
and the other around the higher $\Delta m^2$ corresponding to the 
second minima (high-LMA).
The former is preferred by the \kl data and to a 
greater extent by the global solar data.
At 99.73\% the demarkation between the two zones disappear.

In Table 2 we show the allowed ranges of the values of the parameters 
at 99\% C.L. obtained from the global analysis including the 
solar and \kl  spectrum data. 
The allowed range of $\tan^2\theta$ is not reduced 
in the preferred low-LMA zone with the inclusion of KamLAND data, 
although in the high-LMA zone it is somewhat restricted. 

\begin{table}
\begin{center}
\begin{tabular}{cccc}
\hline\hline
Allowed &99\% C.L. Range of &
99\% C.L. Range of \\
Zone& $\Delta m^2$ in eV$^2$&$\tan^2\theta$ \\
\hline\hline
low-LMA &
$5.3 \times 10^{-5}< \Delta m^2 <9.9 \times 10^{-5}$
& $ 0.28 < \tan^2\theta <0.79$\\
high-LMA &
$1.3 \times 10^{-4}< \Delta m^2 <1.8 \times 10^{-4} $
& $ 0.34 < \tan^2\theta <0.55$\\
\hline\hline
\end{tabular}
\caption
{Range of parameter values allowed at 99\% C.L. from the global solar and 
\kl spectrum analysis. 
}
\end{center}
\end{table}

For the LOW solution we get 
$\chi^2_{min}=97.28$. 
This implies that LOW is 
now ruled out at 4.4$\sigma$ w.r.t. the LMA solution.
The maximal mixing solution
($\chi^2_{min}=88.89$) is 
disfavored at 3.4$\sigma$ (w.r.t. LMA) from the combined \kl+ solar analysis.

\section{Projected Analysis}

\begin{figure}[ht]
\centerline{\psfig{figure= 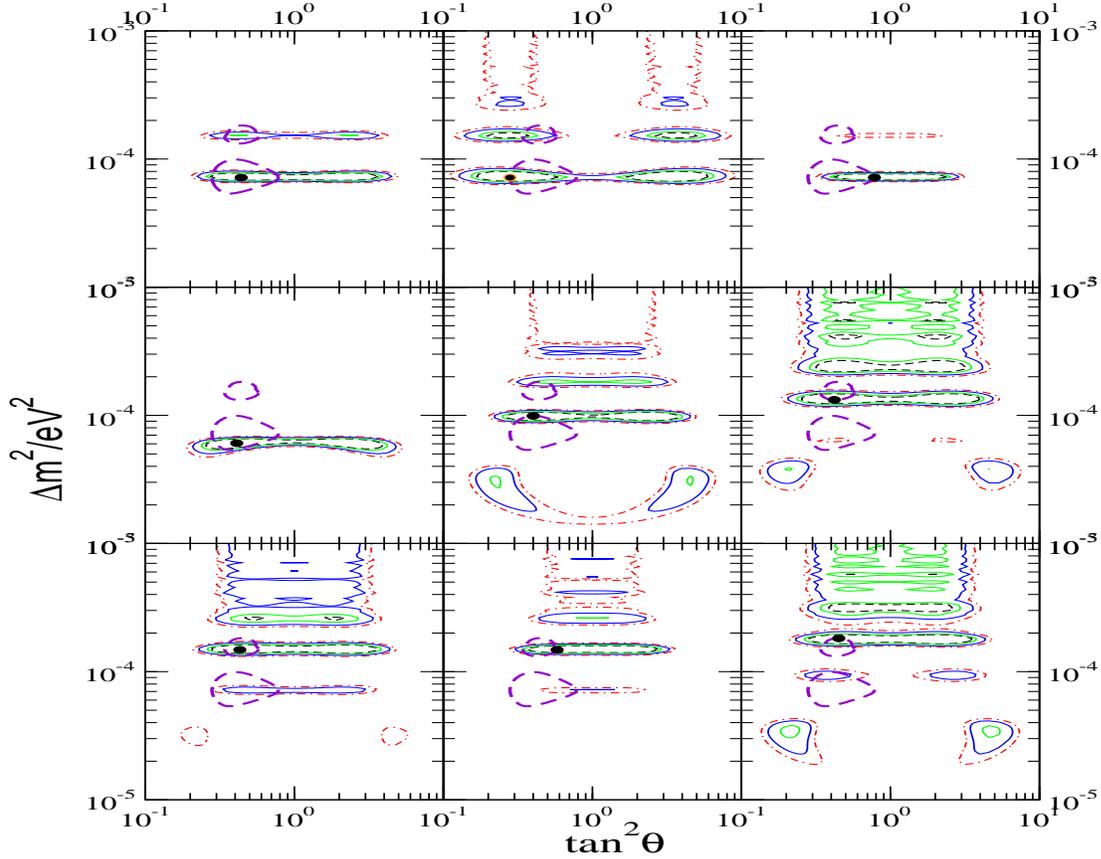 ,height=5.0in,width=6.0in}}
\caption{The 90\%, 95\%, 99\% and 99.73\% C.L. contours for
using the 1 kton-yr 
projected KamLAND spectrum.
The different panels are for the simulated spectrum at
values of $\Delta m^2$ and $\tan^2\theta$ indicated by the 
black dots.
The dashed line shows the currently allowed 99\% C.L. contour 
from solar+KamLAND analysis. 
}
\label{kamonlyspec}
\end{figure}

\begin{figure}[ht]
\centerline{\psfig{figure=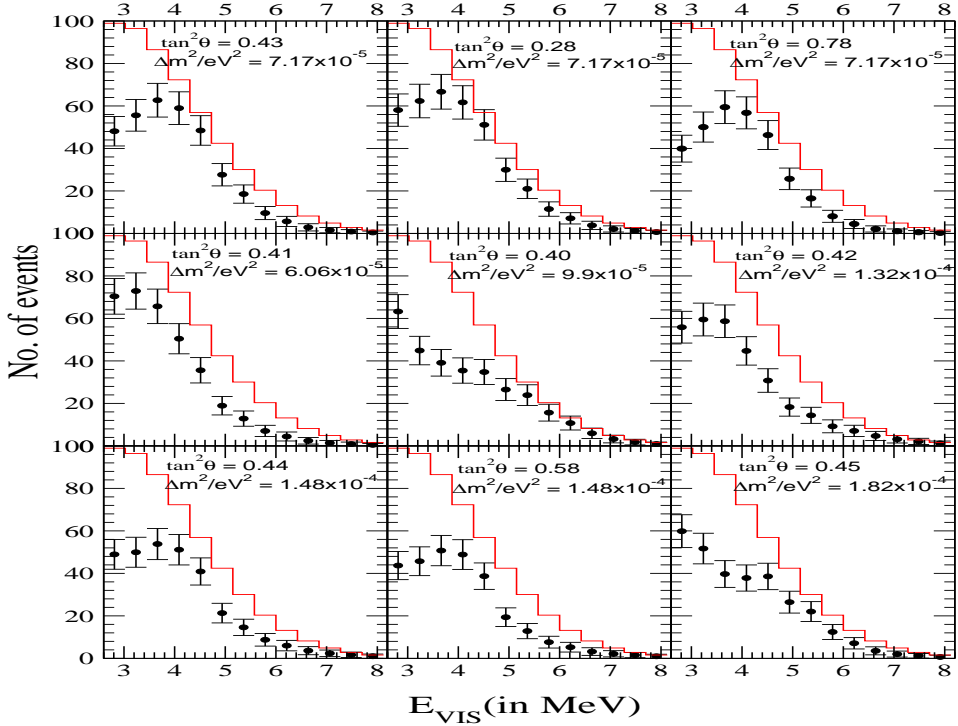,height=4.5in,width=5.5in}}
\caption{The 1 kton-yr simulated KamLAND spectrum for the different sets
of $\Delta m^2$ and $\tan^2\theta$ corresponding to Figures \ref{kamonlyspec}
and 
\ref{kamspec}. The histogram shows the unoscillated spectrum for 1 kton-yr.}
\label{spectrum}
\end{figure}
\begin{figure}[ht]
\centerline{\psfig{figure=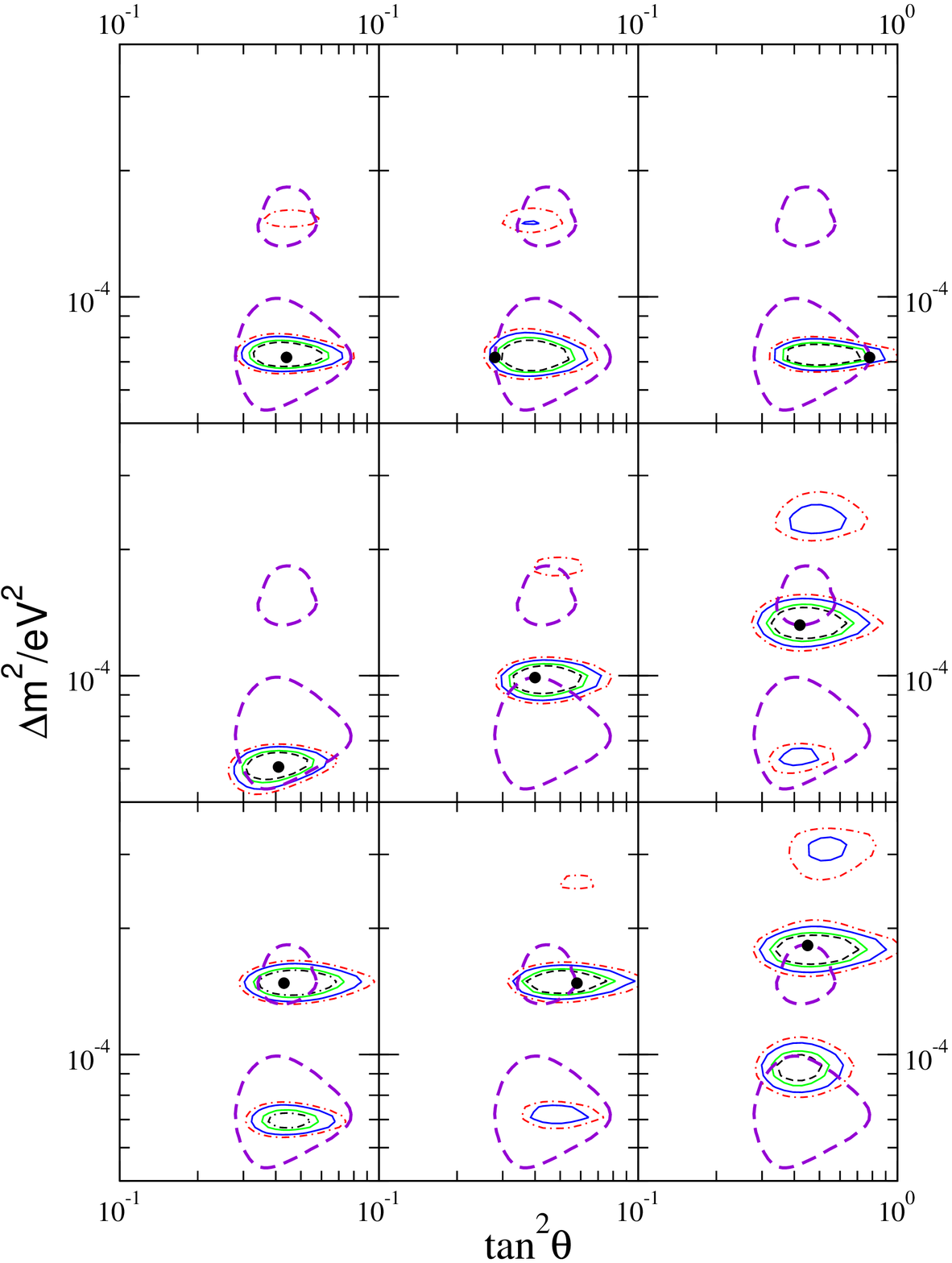,height=5.0in,width=6.0in}}
\caption{Same as in Figure. \ref{kamonlyspec} but from a
combined analysis using the solar+CHOOZ+1 kton-yr
projected KamLAND spectrum data.
Since the solar data disfavour the 
$\theta > \pi/4$ regions this figure is plotted with $\tan^2\theta$ 
extending upto  1.0. 
}
\label{kamspec}
\end{figure}

In this section we explore whether the future
\kl 
spectrum data would be able to determine the allowed zones more precisely. 
In particular, it would be interesting 
to see whether \kl,
either by itself or in conjunction with the solar data, 
can choose between one of the two allowed islands, and 
pin down the values of the mass and mixing parameters unambiguously.
We try to look into this by doing
a statistical analysis of the projected 1 kton-yr \kl data.
We choose  sample
values of $\Delta m^2$ and $\tan^2\theta$, 
from the allowed zones obtained from the global solar+KamLAND  
analysis (cf. fig.\ref{klrtspecsol}) 
and simulate the spectrum at these points 
for 1 kton-yr of data (approximately 2.5 years of \kl live-time ), 
using a randomizing procedure 
which takes care of the 
fluctuations. 
We use these simulated spectra in a $\chi^2$ analysis and 
reconstruct the allowed regions in the $\Delta m^2$--$\tan^2\theta$
plane.
In Figure \ref{kamonlyspec} we display these reconstructed
allowed regions from 
1 kton-yr projected \kl spectrum alone.
Figure \ref{spectrum} shows the 1 kton-yr simulated spectrum with
errorbars at the
$\Delta m^2$ and $\tan^2\theta$ corresponding to each of the panels
of Figure \ref{kamonlyspec}.
In Figure \ref{kamspec} we present the allowed regions from a combined
analysis of global solar data and 1 kton-yr \kl spectrum data, simulated at the
same set of points as  in the previous two figures. The dashed lines 
in Figures  \ref{kamonlyspec} and \ref{kamspec} give the 
current 99\% C.L. allowed regions.
For the analysis with the projected \kl spectra we assume
Gaussian statistics, which is more appropriate in this case.

First let us make a note on the rationale of the representative values of 
$\Delta m^2$ and $\tan^2\theta$ chosen to simulate the spectrum. 
Panel 1 corresponds to the \kl spectrum simulated at the low-LMA best-fit 
while panel 7 corresponds to that generated at the high-LMA best-fit. 
These are the two favored points from the current data. 
We also simulate the 1 kton-yr \kl spectrum 
at few other points 
deviated from the best-fits. 
In  panel 2(3)  we have chosen $\Delta m^2$ corresponding to the low-LMA
best-fit but a lower(higher) value of $\theta$. 
Similarly panel 8 is for $\Delta m^2$ corresponding to
the high-LMA best fit
but at a higher value of $\theta$.
The panels with same $\Delta m^2$ but different $\theta$ are chosen to
demonstrate the impact of $\theta$ on the
reconstructed regions. We choose $\Delta m^2$ values lower(higher) than the
low(high)-LMA best-fit in  panels 4(5).
Panels 6 and 8 display the reconstructed regions 
for $\Delta m^2$ values lower and higher than the high-LMA best-fit 
respectively. 
These set of values give adequate coverage
for studying the 
projected sensitivity of the reconstructed 
regions on the choice of
$\Delta m^2$ and $\theta$  
currently allowed at 99\% C.L..   

Figure \ref{kamonlyspec} addresses the issue if future KamLAND data can by 
itself make a sharper demarkation between the two allowed 
islands and the dependence of this on the oscillation 
parameters. 
The  figure shows that 
if we choose the simulation point at  higher $\Delta m^2$ 
values (panels 5, 6, 7, 8 and 9) then there are large allowed regions 
from the spectrum data, extending upto $10^{-3}$ eV$^2$. 
The Figure \ref{spectrum} shows that for these values of $\Delta m^2$ the 
spectral suppression tends to become flat (undistorted) 
leading to increased fuzziness. 
The same is also true for lower values of $\theta$. 
The lesser distortion at low values of $\tan^2\theta$ is due to a 
diminished energy dependence resulting from a small oscillatory term
in the survival probability \cite{raj}.
This also leads to an increase in allowed area (cf. panel 2) although 
to a lesser extent as compared to the panels with high $\Delta m^2$ 
values discussed earlier. 
There are also some allowed regions at $\Delta m^2$  
values  lower than that allowed by the current 99\% C.L. contour
in panels 5, 6, 7 and 9.  
However the figure reveals that the allowed $\Delta m^2$ range around the 
low-LMA and high-LMA zones get reduced in most of the panels.  
With 
1 kton-yr \kl spectrum data the region around 
$\Delta m^2 \sim 10^{-4}$ eV$^2$  present in Figure 
\ref{klrtspecsol}
gets disallowed in all the panels and the 
low-LMA and high-LMA regions 
get bifurcated
even at 3$\sigma$ level. 
In general, tighter 
constraints in the Figure \ref{kamonlyspec} are associated with spectra 
with a shape significantly different from the no oscillation spectrum.
For the KamLAND baselines this 
corresponds 
to a lower $\Delta m^2$ than 
the present best-fit. 
Panel 4, which is at the global solar best-fit, shows that 
for such cases just the 
1 kton-yr spectrum data from KamLAND can pick out a sharply defined 
allowed zone around the 
simulation point unambiguously. 

In Figure \ref{kamspec} we show the C.L. allowed regions from a
combined analysis of the global solar + CHOOZ + 1 kton-yr
KamLAND simulated spectrum data. Through these plots we
investigate the impact of the 
current solar data to resolve the ambiguity still admitted by the 
1 kton-yr projected \kl spectrum data.
We find that the solar data is instrumental in ruling out a large 
part of the parameter space allowed by the 1 kton-yr \kl
only analysis.
A comparison of the panels 1, 2, 3 and 5  in the figures 
\ref{kamonlyspec} and \ref{kamspec}
shows that  for spectrum simulated in the low-LMA region 
the inclusion of the solar data 
reduces the statistical significance of the 
high-LMA zone.  
It also disallows the high $\Delta m^2$ regions above $2\times 10^{-4}$ 
eV$^2$. 
In these regions the solar data 
requires a $^8{B}$ flux normalisation factor ($f_B$) $\sim$ 0.8 which is 
in conflict with the SNO NC measurement of $f_B \sim 1.0$, thus 
disfavoring these zones.  
For spectrum simulated in the high-LMA region
in panels 6 to 9, even though 
the large allowed regions beyond $2\times 10 ^{-4}$ eV$^2$ 
as well as the allowed regions at low $\Delta m^2$
get mostly removed by the solar 
data, 
the ambiguity between the allowed islands remain. 
In fact a comparison with Figure \ref{kamonlyspec} reveals that 
for these panels the inclusion of the 
solar data increases the statistical significance 
of the low-LMA allowed regions. 
The \kl spectral data allows the low-LMA region
only at 99\% C.L. in panels 7, 8 and 9, and at 99.73\% C.L. in panel 6. 
However with the inclusion of the solar data in the analysis, 
these regions become allowed at 90\% and 99\% C.L. respectively,
as the solar data prefers the low-LMA zone. 
Therefore to remove the ambiguity for the spectrum corresponding to 
the high-LMA 
zones, one would require 
KamLAND data with higher statistics, which will be 
able to determine the spectral shape 
and 
hence $\Delta m^2$ more precisely.

\section{Conclusions}

In conclusion, we have investigated the impact of the 
first results from KamLAND on neutrino mass and mixing parameters
in conjunction with the global solar neutrino data. 
KamLAND is completely consistent with the LMA solution, to the extent 
that the observed KamLAND rate is close to that predicted by the 
best-fit point of the LMA solution 
to the solar neutrino problem. As a result  
the combined analysis of the solar and KamLAND rates data 
allows a large area within the solar LMA region and the 
global solar best-fit does not change much with inclusion of 
\kl rates.
The constraining capabilities of the spectrum data 
is  much stronger 
and with only 145 days observed  spectrum  
\kl can
exclude certain parts of the LMA parameter space. 
After including the spectral data 
the  
allowed  LMA zone  consists mainly of two disconnected regions, one 
around the best-fit and another at a higher $\Delta m^2$. The 
two zones merge at $3\sigma$.  
Maximal mixing though allowed by the \kl alone, is found to be 
still disfavored by the combined solar and \kl data at more than 
$3\sigma$. The LOW solution which was allowed at $3\sigma$ from the 
global solar data and which predicts null oscillations 
in \kl is now disfavored at almost $5\sigma$ w.r.t the LMA solution. 

With LMA now confirmed, the next focus of KamLAND would be a more
accurate determination 
of the mass parameter 
by distinguishing between the two allowed sectors in the LMA
region.
We have explored this  through a projected analysis with 1 ktyr 
simulated data. 
With 1 kton-yr projected spectrum data the allowed $\Delta m^2$ 
ranges around both 
low-LMA and high-LMA zones decrease in size and they get separated 
at 3$\sigma$ by the spectrum data itself. 
The inclusion of the solar data disfavours(favours) the high(low)-LMA zone 
if the 
spectrum is simulated in the low(high)-LMA area.
Thus 
the allowed areas become more precise for low-LMA spectrum 
while ambiguity 
between the two zones remains for high-LMA  spectrum. 
A higher statistics from \kl
is expected to resolve this ambiguity. 
A more precise
determination of the mixing angle will be possible from a more accurate
measurement of the CC/NC ratio at SNO.

\vskip 10pt
{\bf  Acknowledgment }
We acknowledge S. Pakvasa for useful discussions. 
RG would like to thank John Beacom for a helpful discussion.
SC acknowledges discussions with S.T. Petcov.


\end{document}